\documentclass[12pt,preprint]{aastex}

\newcommand{\HI}{H~{\sc i}} 

\newcommand{\kms}{${\rm km~s^{-1}}$}

\slugcomment{{\em To appear in ApJS}}

\shortauthors{MCCLURE-GRIFFITHS ET AL} 
\shorttitle{THE SGPS: \HI\ OBSERVATIONS AND ANALYSIS}

\begin{document} 

\title{The Southern Galactic Plane Survey: \HI\ Observations and Analysis}

\author{N.\ M.\ McClure-Griffiths,\altaffilmark{1} John M.\ Dickey,\altaffilmark
{2,3} B.\ M.\ Gaensler,\altaffilmark{4} A.\ J.\ Green,\altaffilmark{5}
Marijke Haverkorn,\altaffilmark{4} and S.\ Strasser\altaffilmark{2}}

\altaffiltext{1}{Australia Telescope National Facility, CSIRO, PO Box 76, Epping
 NSW 1710, Australia; naomi.mcclure-griffiths@csiro.au} 
\altaffiltext{2}{Department of Astronomy, University of Minnesota, 116 Church St
 SE, Minneapolis, MN 55455; john@astro.umn.edu, strasser@astro.umn.edu}
\altaffiltext{3}{Current address: School of Mathematics and Physics,
 University of Tasmania, Private Bag 21, Hobart TAS 7001, Australia}
\altaffiltext{4}{Harvard-Smithsonian Center for Astrophysics, 60 Garden Street M
S-6, Cambridge, MA 02138; bgaensler@cfa.harvard.edu, mhaverkorn@cfa.harvard.edu}
\altaffiltext{5}{School of Physics, University of Sydney, NSW 2006, Australia; agreen@physics.usyd.edu.au}

\authoraddr{Address correspondence regarding this manuscript to: 
                N. M. McClure-Griffiths
                ATNF, CSIRO
                PO Box 76
		Epping NSW 1710
		Australia }
\begin{abstract}
We describe the \HI\ component of the Southern Galactic Plane Survey
(SGPS).  The SGPS is a large-scale project to image at arcminute
resolution the \HI\ spectral line and 21 cm continuum emission in
parts of the plane of the Milky Way.  The survey covers Galactic
longitudes $253\arcdeg \leq l \leq 358\arcdeg$ and latitudes $|b|
\leq1\fdg5$ (SGPS I), plus a first quadrant extension covering
$5\arcdeg \leq l \leq 20\arcdeg$ and $|b| \leq 1\fdg5$ (SGPS II).  The
survey combines data from the Australia Telescope Compact Array and
the Parkes Radio Telescope for sensitivity to angular scales ranging
from 2 arcminutes to several degrees.  The combined data cover 325
${\rm deg^2}$ and have an rms sensitivity of $1.6$ K.  Here we
describe the \HI\ observations and data reduction in detail, and present
examples from the final data products. The complete dataset is
publicly available through the Australia Telescope National Facility's
\HI\ Surveys archive.  This dataset provides an unprecedented view of
the neutral component of interstellar hydrogen in the inner Milky Way.
\end{abstract}

\section{Introduction}
\label{sec:intro}
The physics of the interstellar medium (ISM) is diverse and
complicated.  The cycle of stellar formation and death takes the ISM
through a wide range of physical conditions, a number of which can be
traced by the 21-cm hyperfine emission line produced by neutral
hydrogen (\HI).  Tracing both structure and dynamics, \HI\ is observed
in emission and absorption, each probing different thermal states of
the gas.  Recent observations by \citet{heiles03b} and
\citet{dickey03} show that \HI\ observed in absorption traces gas with
temperatures as cold as 15 K, whereas \HI\ emission observations
generally trace gas with temperatures in the range of few thousands
\citep{dwarakanath02}.  \HI\ densities vary over a similarly broad
range, from $\sim 10^{-2}~{\rm cm^{-3}}$ in the diffuse warm neutral
medium to $\sim 10^2~{\rm cm^{-3}}$ in regions of the cold neutral
medium.  The Milky Way provides an ideal laboratory in which to study
the neutral ISM at sensitivities and spatial scales that are
unattainable in external galaxies.  Despite many years of observations
of the ISM, our understanding of the medium is still simplistic,
relying heavily on a three-phase model \citep[e.g.][]{mckee77}.  We
have yet to adequately model the variety of temperatures, densities
and structures known to exist in the ISM.  Our lack of understanding
is not only from the theoretical standpoint, but also from the lack of
observational data probing \HI\ physics over a wide range of spatial
scales.  It is clear that there is a great deal to learn about the
structure, dynamics and thermal state of the neutral ISM from high
resolution surveys of Galactic \HI.

To reach this goal and develop a better understanding of the Galactic
ISM the International Galactic Plane Survey (IGPS) is systematically
imaging \HI\ in the disk of the Milky Way.  The IGPS consists of
three individual surveys, the Canadian Galactic Plane Survey
\citep[CGPS;][]{taylor03}, the VLA Galactic Plane Survey
\citep[VGPS;][]{taylor02}, and the Southern Galactic Plane Survey
(SGPS, described here).  These surveys excel because of their
exceptional spatial dynamic range, combining single dish and
interferometer data for arcminute resolution imaging over areas of
many degrees.

The Southern Galactic Plane Survey \citep{mcgriff01a} is a survey of
the 21-cm continuum and \HI\ spectral line emission in the fourth and
parts of the first and third quadrants of the disk of the Milky Way.
The survey provides a complete \HI\ and 21 cm full polarization
(Stokes I, Q, U, \& V) dataset of the area $253\arcdeg \leq l \leq
358\arcdeg$, $b = \pm 1\fdg5$ (SGPS I) and an extension into the first
quadrant covering $5\arcdeg\leq l \leq 20\arcdeg$, $b = \pm 1\fdg5$
(SGPS II).  The SGPS is comprised of two surveys: a low resolution
(FWHM = 15\arcmin) survey completed with the Parkes 64m radio
telescope and a high resolution (FWHM = 2\arcmin) interferometric
survey completed with the Australia Telescope Compact Array (ATCA).
The data are combined to provide a dataset that is sensitive to more
than two orders of magnitude in spatial scales.  The combined data
have an angular resolution of $\sim 2$ arcmin, spectral resolution of
$0.8$~\kms\ and a sensitivity limit of $\sim 1.6$ K in the line.  The
main SGPS parameters are given in Table~\ref{tab:params}.

The scientific objectives of the SGPS are to provide a high-resolution
21cm polarized continuum and \HI\ atlas of the Galactic plane with
which to study the distribution and dynamics of both the warm and cool
components of the ISM over size scales ranging from parsecs to
kiloparsecs \citep{mcgriff01a,mcgriffphd}.  Scientific results from
the survey and its ``test region'' have been presented in a number of
papers. An overview of the \HI\ emission and absorption in the test
region was published in \citet{mcgriff01a}.  Other SGPS papers
focusing on the test region data include an analysis of the \HI\
emission spatial power spectrum \citep{dickey01}; a discussion of the
abundance and temperature of cool \HI\ clouds in the inner Galaxy
\citep{dickey03}; an examination of extended polarized emission
structure and rotation measures of point sources \citep{gaensler01};
and an analysis of rotation measure fluctuations indicating an outer
scale for turbulence of $\sim 2$ pc \citep{haverkorn04}.  Using the
SGPS as a whole we have discovered a new, outer spiral arm in the
fourth quadrant of the Milky Way \citep{mcgriff04}; probed the physics
of individual \HI\ supershells
\citep{mcgriff00,mcgriff01c,mcgriff03b}; discussed the Galactic
distribution of \HI\ supershells from a catalog of new supershells
discovered in the SGPS \citep{mcgriff02a}; and explored the properties
of \HI\ self-absorption clouds discovered in the SGPS
\citep{kavars03,kavars04}.

Here we describe the SGPS in detail, focusing on the observing
strategy and data reduction techniques.  In particular, we describe
the \HI\ data; the continuum and polarization data require vastly
different reduction techniques and will be described in a separate
paper (M.\ Haverkorn et al.\ 2004, in preparation).  This paper is
organized as follows: in \S \ref{sec:parkes} we describe the
low-resolution Parkes multibeam survey, in particular the data
reduction strategies developed explicitly for this project.  In \S
\ref{sec:atca} we describe the ATCA observations and data reduction,
especially where the reduction deviates from standard techniques.  The
combination of our single dish and interferometer data is described in
\S \ref{sec:combination}.  Finally, the complete data products are
described in \S \ref{sec:products} and representative images from the
combined data cubes are presented in \S \ref{sec:images}.  The SGPS
data are publicly available from the Australia Telescope National
Facility (ATNF) \HI\ Surveys
archive.\footnote{http://www.atnf.csiro.au/research/HI/common/}
\section{Parkes Survey}
\label{sec:parkes}
The low resolution portion of the SGPS was observed with the Parkes
Radio Telescope.  This survey covers the area $253\arcdeg \leq l \leq
20\arcdeg$, $b=\pm10\arcdeg$ with an angular resolution of
15\arcmin. The Parkes survey was primarily designed to provide
information about large scale structure not sampled in the ATCA
survey.  Expanding the latitude coverage of the Parkes survey to trace
large-scale structures away from the Galactic plane provides
additional information about the gas in a thicker disk and has led to
the discovery of many new large \HI\ shell structures \citep{mcgriff02a}.
Therefore, the Parkes survey is released as a separate data product.
Here we describe the observational and data reduction techniques used
for this survey.

\subsection{Observations}
Single dish observations were made with the 21cm multibeam system at
prime focus on the Parkes 64m Radio Telescope near Parkes, NSW,
Australia.  This is a focal plane array with thirteen independent
feeds, each with dual, cryogenically cooled, orthogonal linear
polarization receivers with stable, low system temperatures
($T_{sys}\sim 25$ K).  The feeds are packed in a hexagonal pattern
with a separation of $29\farcm1$ between the centers of adjacent feeds
\citep{staveley-smith96}.  At $\lambda = 21$ cm the FWHM of the Parkes
beam is $14\farcm4$, so the multibeam array does not fully sample the
focal plane.  For this project only the inner seven beams were used.

Data were obtained by mapping ``on-the-fly'', driving the telescope at
a constant Galactic longitude through 3 degrees of Galactic latitude
at a rate of $1.1~{\rm deg~min^{-1}}$.  Data samples were written to
disk every 5s.  In order to maximize sky coverage and reduce redundant
samples the multibeam receiver platform was rotated to maintain an
angle of $19\farcm1$ between the face of the hexagon and the scan
direction.  The platform rotation results in seven parallel beam
tracks, each separated by $9\farcm5$.  To ensure Nyquist sampling,
interleaved scans were made by offsetting the receiver package by 2.5
beam spacings and scanning in the opposite direction.  The offset
scans reduced the spacing between adjacent tracks to $4\farcm7$ and
ensured that independent feeds are responsible for adjacent tracks.
Along the scan direction the data samples are spaced by $\sim
5\farcm5$, which smears the beam slightly in the scan direction.  In
order to reduce the effects of variations in the system temperature
with elevation, we made an effort to scan the telescope at a nearly
constant zenith angle of approximately 30\arcdeg.

Observations were conducted during five sessions between December 1998
and November 2000.  The first set of observations on 15 and 16
December 1998 covered the ATCA survey region of $253\arcdeg \leq l
\leq 358\arcdeg$, $|b| \leq 1\fdg5$.  We conducted repeat observations
of this same region at three month intervals for a year in order to
reduce the annual variations in \HI\ spectra due to stray radiation
\citep{kalberla80} and to improve the sensitivity of the observations.
Ultimately the ATCA survey region was observed 10 times.  During
subsequent observing runs in June 1999, September 1999, and March 2000
the latitude range was extended to cover $|b| \leq 10\arcdeg$.
Finally, between 28 October and 5 November 2000 the survey was
extended through the Galactic Center to $l=20\arcdeg$.

The data were recorded in frequency switched mode, switching in band
between a center frequency of 1419.0 and 1422.125 MHz with every 5s
cycle.  We used the narrow-band multibeam correlator \citep{haynes99},
with a maximum bandwidth for each frequency of 8 MHz, with 2048
channels, giving a channel width of 3.9 kHz ($\Delta v = 0.82$ \kms).
After bandpass correction using frequency switched samples the usable
bandwidth was approximately 4.5 MHz.

\subsection{Bandpass and Brightness Temperature Calibration}
\label{subsec:calibration}
The IAU standard narrow line calibration regions S6 and S9 were
observed daily for absolute brightness temperature and bandpass
calibration.  For these observations each beam of the array was
pointed on source for a 60s integration.  These regions were also
observed near a zenith angle of 30\arcdeg\ so as not to compromise
calibration with a significantly different gain and system temperature
from the survey fields.

The bandpass shape is relatively stable over periods of days.
However, there are variations in the front-end gain versus frequency
shape on the timescale of seconds to minutes that can largely be
removed by frequency switching.  To remove the bandpass ripples, each
spectrum was divided by the previous frequency switched sample to
create a quotient spectrum.  After this first reduction step, the
shape of the quotient spectrum is stable to the 1-3\% level over the
course of a day.  Next, the residual bandpass shape was corrected by
dividing the quotient spectrum by a bandpass template.  The bandpass
template was determined daily for each beam and polarization from a
time-averaged quotient spectrum on one of the standard line regions,
S6 or S9.  These spectra were normalized to a mean of one and then
hand-fit by a series of Fourier components to create a bandpass
template, $f (\nu)$.  In general, the narrower line width of S6
($\Delta v = 10.6$ \kms\ for S6, versus $\Delta v = 15.8$ \kms\ for
S9) resulted in a better fit for the template.

In standard processing of \HI\ single dish spectra the antenna
temperature is calculated as $T_a = T^{\prime}_{\rm sys} (T_{obs} -
T_{ref}) T_{ref}^{-1}$, where $T^{\prime}_{\rm sys}$ is the system
temperature determined online during the integration, $T_{obs}$ is the
observed spectrum and $T_{ref}$ is the subsequent frequency-switched
spectrum.  The application of this technique, with some polynomial
bandpass fitting, is relatively successful at removing ripples from
the bandpass shape, but it also removes any continuum emission.  For
the SGPS we are interested in the continuum emission as well as the
\HI\ line so we employed a non-standard technique.  For each spectrum
we calculated the mean value, $T_{bas}$, of the reference spectrum
over the area of the line, which should be the sum of the sky
continuum temperature and the actual system temperature, $T_{\rm
sys}$.  For $T_{\rm sys}$ we used an average system temperature
calculated on the standard line regions, which we apply to an entire
day's observations.  This is a generally valid assumption because the
system temperature of the multibeam is extremely stable. The bandpass
corrected spectra were then calculated as:
\begin{equation}
T_a (\nu) = T_{bas} \,
\frac{T_{obs}(\nu)}{f(\nu)\,T^{\prime}_{ref}(\nu)} - T_{sys}\,\,,
\label{eq:calib}
\end{equation}
where $T^{\prime}_{ref}(\nu)$ is a smoothed version of the frequency
switched, or reference, spectrum.  This technique differs from the
standard single dish \HI\ bandpass calibration because it only
subtracts the absolute system temperature, not the sum of the system
temperature and the continuum emission.  The resultant spectra contain
the observed continuum emission and have baselines that are typically
flat to less than 1K.  They are limited, however, by the accuracy in
our determination of $T_{\rm sys}$.  The value of $T_{\rm sys}$
contains not only the receiving system temperature, but also the
ground radiation temperature, which is elevation dependent and poorly
known at 21 cm the for Parkes telescope.  Because of the difficulty in
estimating total power continuum levels with a single dish, $T_{\rm
sys}$ is only accurate to $\sim 10 - 15$\%.  Fortunately, for data
products from which the continuum emission is subtracted later in the
analysis this problem is alleviated.
 
After bandpass correction, the integral over part of the line on the
standard regions was used to determine a calibration scale factor,
$C$, to convert antenna temperature to absolute brightness
temperature, $T_b = C \,T_a$.  The calibration factor was calculated
for each beam and each polarization by comparing the measured line
integral on the standard line regions to the published
\citet{williams73} values for the line integral, such that
$C=I_{true}/I_{obs}$.  For S6 the line integral was calculated over
the velocity range $-5.8 ~{\rm km~s^{-1}} \leq v \leq +4.8~{\rm
km~s^{-1}}$ and is $I_{true} = 299~{\rm K~km~s^{-1}}$.  The S9 line
integral over the velocity range $+1.05 ~{\rm km~s^{-1}} \leq v \leq
+14.75~{\rm km~s^{-1}}$ is $I_{true} = 953~{\rm K~km~s^{-1}}$.  Using
the line integral to scale the measured antenna temperature we find
that the peak brightness temperature for S6 is $T_{max} = 57 \pm 2$ K
and $T_{max} = 90\pm3$ K for S9.  For comparison, another recent
Parkes multibeam survey with a similar observational setup found
$T_{max} = 83$ K for S9 \citep{bruens05}.  The discrepancy may be due
to the difference between the velocity range used for the line
integral in \citet{williams73}, from which our calibration was derived
and \citet{kalberla82}, from which the \citet{bruens05} calibration
was derived.

Finally, the spectra are shifted to the Local Standard of Rest (LSR)
reference frame.  The correction is calculated using standard
techniques and is applied as a phase shift in the Fourier domain.

\subsection{Imaging}
The fully calibrated and velocity shifted Parkes spectra were imaged
using {\em Gridzilla}, a gridding tool created for use with Parkes
multibeam data and found in the ATNF subset of the {\em aips++} (now
{\em casa}) package.  The gridding algorithm is described in detail by
Barnes et al.\ (2001)\nocite{barnes01}.  The SGPS data were gridded
using a weighted median technique with a cellsize of 4\arcmin, a
Gaussian smoothing kernel of 16\arcmin, and a smoothing kernel cutoff
radius of 10\arcmin. The Gaussian smoothing kernel and cutoff radius
reduce the effective angular resolution of the gridded data to $\sim
16\arcmin$, which is slightly larger than the 14\farcm4 FWHM of the
Parkes beam at 1420 MHz. The effective resolution is estimated from
Gaussian fits to point sources in the Parkes continuum images and to a
beam map made on the calibration source PKS B1934-638.  The imaged
data were imported into the MIRIAD data reduction package
\citep{sault04} for further analysis and continuum subtraction. Image
domain continuum subtraction was performed using a first order
polynomial fit to the off-line channels.

\section{Australia Telescope Compact Array Survey}
\label{sec:atca}
The high resolution portion of the SGPS was completed with the
Australia Telescope Compact Array (ATCA) synthesis imaging telescope.
The ATCA survey is divided into two parts: SGPS I, covering the area
$253\arcdeg \leq l \leq 358\arcdeg$, $|b| \leq 1\fdg5$ and SGPS II,
covering the area $5\arcdeg \leq l \leq 20\arcdeg$, $|b| \leq 1\fdg5$.
The Galactic center area from $358\arcdeg \leq l \leq 5\arcdeg$ has
also been observed, but will be presented separately because it has a
different observing and reduction strategy. Here we describe the
observational and data reduction strategies used for both SGPS I and
II.

\subsection{Observations}
The observations for SGPS I were conducted between December 1998 and
August 2000, with re-observation of several areas in June 2001.  SGPS
II was observed between June and October 2002.  The observational
strategies are slightly different for the two data sets.

The ATCA is a six element interferometer near Narrabri, New South
Wales, Australia.  It consists of five 22 m antennas that can be
positioned along on a 3 km east-west track or a perpendicular 200 m
north-south track and a sixth antenna fixed a further 3 km west of the
end of the east-west track, for a maximum baseline of 6 km.  The
antennas can be arranged in a number of configurations.  The SGPS
takes advantage of the ATCA's most compact configurations with maximum
baselines (excluding the 6 km antenna) between 210 m and 750 m to
maximize sensitivity to large scale structures.  The array
configurations used for SGPS I were: 210, 375, 750A, 750B, 750C, and
750D.  The array configurations were chosen to ensure optimum sampling
of the inner {\em u-v} plane and minimize redundant baselines.  All
baselines, in intervals of 15.5 m, from 31 m to 291 m, except 199m,
are sampled.  The long baselines of the 750-m arrays are poorly
sampled and limit our practical resolution to $\sim 2\arcmin$.  Before
phase II of the SGPS the ATCA was upgraded to include a short
north-south track and to add new stations on the east-west track that
improve the coverage of the compact configurations.  Because the SGPS
II extends as far north as $\delta = -11\arcdeg$, the new north-south
baselines were utilized to improve the {\em u-v} coverage and to
reduce the effects of shadowing at low-elevations.  The SGPS II was
observed with the new EW352 and EW367 configurations, which give
nearly complete coverage to baselines of 352 m in intervals of 15.5 m,
and the north-south/east-west hybrid arrays, H75 and H168.  Even with
the hybrid arrays, the synthesized beam for the more northerly of the
SGPS II fields is quite elliptical ($200\arcsec \times 130\arcsec$).

\begin{figure}
\centering
\includegraphics[angle=-90,width={0.6\textwidth}]{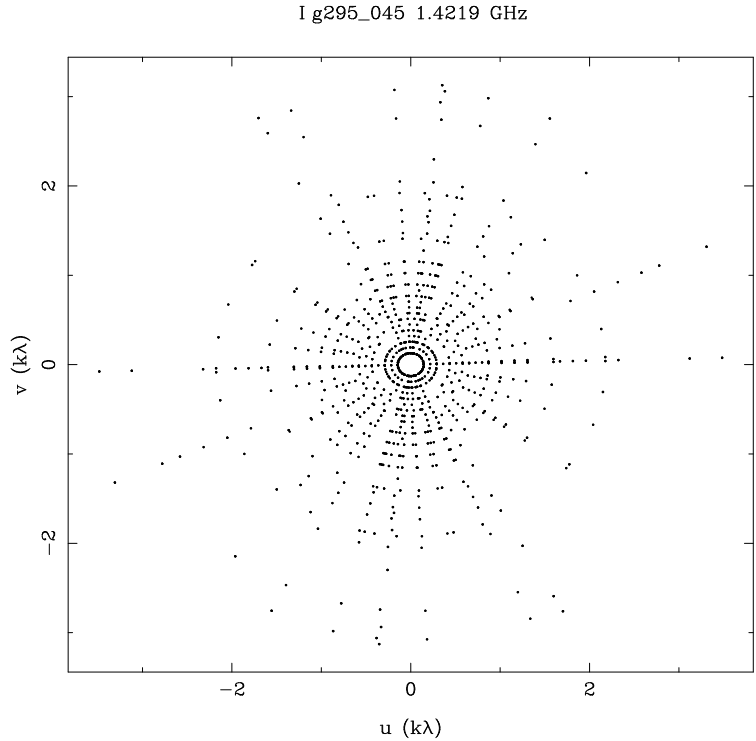}
\caption[]{The {\em u-v} coverage for a typical SGPS I pointing in the
  field g295.  The 6 km antenna has been omitted from this plot to show the coverage of
  the inner {\em u-v} plane.  These data use only the east-west configurations of the ATCA.
\label{fig:uvcover}}
\end{figure}

\begin{figure}
\centering
\includegraphics[angle=-90,width={0.6\textwidth}]{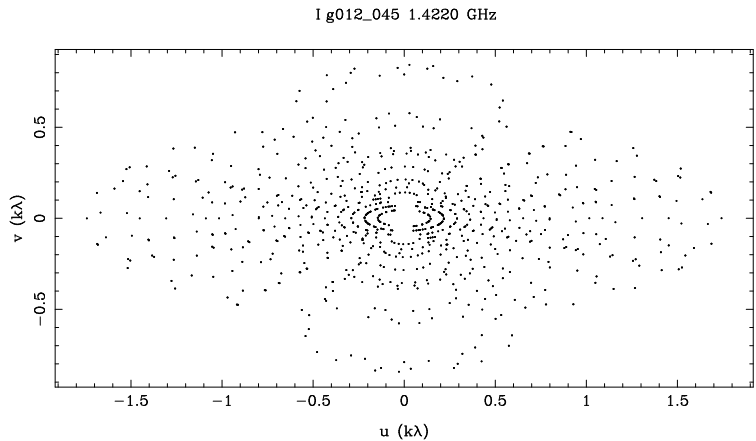}
\caption[]{The {\em u-v} coverage for a typical SGPS II pointing in
  the field g012.  The 6 km antenna has been omitted from this plot to
  show the coverage of the inner {\em u-v} plane.  The {\em u-v}
  coverage of this pointing differs from the coverage shown in
  Figure~\ref{fig:uvcover} because of the use of the north-south hybrid
  arrays in SGPS II and the lower declination of the field.
\label{fig:uvcover2}}
\end{figure}

In order to cover an area much larger than the primary beam of the
ATCA antennas ($\Omega_{pb} = \lambda/D=33\arcmin$), the survey was
conducted as a mosaic of many pointings.  The pointings were arranged
in a hexagonal pattern with Nyquist spacing of 19 arcmin between
adjacent pointings.  The theoretical rms sensitivity for Nyquist
sampled pointings is uniform to within $\leq 1$\% across the entire
observed area.  In the SGPS I there are a total of 2212 pointings
comprising twenty fields of 105 pointings and one field of 112
pointings. In the SGPS II there are 315 pointings divided in three
fields of 105 pointings each.  The fields composed of 105 pointings
cover $5\fdg5 \times 3\arcdeg$, centered on a Galactic latitude of
$b=0\arcdeg$.  Each pointing was observed approximately 40 times with
30 second integrations for a total integration time of at least 20
minutes.  The individual 30 second integrations are distributed evenly
with hour angle.  The exact {\em u-v} coverage varies from pointing to
pointing, but is mostly constant within any given field.  An example
of the {\em u-v} coverage for one pointing near the center of the
field at $l=295\arcdeg$ is shown in Figure~\ref{fig:uvcover}.  The
{\em u-v} coverage for this pointing is typical of the SGPS I fields.
Similarly, an example of the {\em u-v} coverage for a pointing near the
center of the field at $l=12\arcdeg$ is shown in
Figure~\ref{fig:uvcover2}.  This coverage is typical of the SGPS II
fields, although the coverage is much more circular for pointings
near $l=7\arcdeg$ and more elliptical for pointings near
$l=17\arcdeg$. The {\em u-v} coverage shown in
Figure~\ref{fig:uvcover2} differs from the coverage shown in
Figure~\ref{fig:uvcover} because of the use of the north-south hybrid
arrays in SGPS II.

The ATCA has linear feeds, receiving two orthogonal linear
polarizations, $X$ and $Y$, and is capable of observing two frequencies
simultaneously.  All data were recorded in a spectral line mode with
1024 channels across a 4 MHz bandwidth centered at 1420 MHz, and
simultaneously a continuum mode with 32 channels across a 128 MHz
bandwidth centered at 1384 MHz.  In continuum mode the full
polarization products, $XX$, $YY$, $XY$, and $YX$ are recorded,
whereas only the autocorrelations $XX$ and $YY$ are recorded in
spectral line mode.  Only the narrowband data are described here.

\subsection{Calibration and Imaging}
Data editing, calibration and imaging were conducted in the MIRIAD
data reduction package using standard techniques \citep{sault04}.  The
primary flux calibrator for the southern hemisphere, PKS B1934-638,
was observed at least once per day and was used for bandpass and
absolute flux calibration of all data, assuming a flux at 1420 MHz of
14.86 Jy \citep{reynolds94}.  Observations of the secondary
calibration sources PKS B0823-500, PKS B1105-680, PKS B1236-684, and
PKS B1833-210 were used to solve for the time varying gain and delay
corrections.  One secondary calibrator was observed approximately
every hour during the course of observations.  The source PKS
B0823-500 was observed for all fields between $l=255\arcdeg$ and
$l=290\arcdeg$.  PKS 1105-680 was observed for all fields from
$l=290$\arcdeg\ to $l=300\arcdeg$, and PKS 1236-684 was observed for
all fields between $l=300\arcdeg$ and $l=320\arcdeg$.  For fields
between $l=320\arcdeg$ and $l=355\arcdeg$ PKS 1934-638 was used as
both a primary and a secondary calibrator.  Finally, PKS B1833-210 was
used as the secondary calibrator for the three fields in SGPS II.  In
all cases, the gain solutions from the nearest secondary were copied
to the SGPS field data.  For fields flanked by two secondary
calibrators the two solutions were merged when copied to the SGPS
field.

The individual pointings of a field were linearly combined and imaged
using a standard grid-and-FFT scheme.  The mosaicing process uses the
joint approach where dirty images of each pointing are linearly
combined and then jointly deconvolved \citep{sault96}.  This approach
can offer a better deconvolution because information away from the
pointing center is included in the deconvolution process.
Additionally, this method takes best advantage of the mosaicing
technique's ability to recover information on angular scales larger
than the scale determined by the minimum baseline separation,
$\lambda/d_{\rm min} \sim 23\arcmin$, where $d_{\rm min}$ is the
minimum baseline.  Jointly imaging and deconvolving the pointings
increases the maximum angular scale imaged to $\lambda /(d_{\rm min} -
D/2) \sim 36\arcmin$, where $D$ is the diameter of a single antenna.

Super-uniform weighting was used for all \HI\ cubes.  Like uniform
weighting, super-uniform weighting minimizes sidelobe levels to
improve the dynamic range and sensitivity to extended structures.
However, uniform weighting reverts to natural weighting if the field
of view is much larger than the primary beam, as is the case for large
mosaics.  Super-uniform weighting decouples the weighting from the
field size and is therefore more successful than uniform weighting on
large mosaics \citep{sault04}.

Two cubes of each imaged region were made: one containing continuum
emission and one for which the continuum had been subtracted.  For the
continuum subtracted \HI\ cubes, the continuum emission was subtracted
from the calibrated {\em u-v} data prior to imaging \citep{sault94}.
These data were imaged without the 6 km antenna because of the poor
sensitivity of the long baselines.  For the continuum subtracted
cubes, the pointings from two adjacent fields were jointly imaged to
produce cubes covering $11\arcdeg \times 3\arcdeg$ ($l\times b$).  For
the \HI\ cubes containing continuum emission the 6 km baselines were
retained for use with \HI\ absorption experiments.  All cubes were
then deconvolved using a maximum entropy deconvolution algorithm
\citep{sault96}.  The continuum-subtracted cubes for SGPS I were
restored with a circular synthesized beam of $130\arcsec$.  Because of
their more northernly declinations, the SGPS II cubes are restored
with slightly elliptical beams, $160\arcsec \times 110\arcsec$ for
field g010 and $200\arcsec \times 130\arcsec$ for field g015.  The
SGPS I and II cubes with continuum emission and the 6 km antenna were
restored with varying beam sizes, given in Table~\ref{tab:contcubes}.

\section{Single Dish and Interferometer Data Combination}
\label{sec:combination}
Although mosaicing recovers information on larger angular scales than
normal interferometric observations by reducing the effective shortest
baseline, the ATCA images alone are not sensitive to angular scales
larger than $\sim 36$ arcmin.  In order to recover information on
scales as large as the image size, we have combined the ATCA data with
the Parkes data of the same region, which continuously sample the {\em
u-v} plane between zero baseline separation and the 64 m diameter of
the dish.  There are a number of methods for combining single dish and
interferometric data.  The merits of the various methods have been
discussed extensively by others, \citep[e.g.][]{stanimirovic02}, so we
will not repeat the discussion.  For the SGPS we combine the data in
the Fourier domain after deconvolution, as implemented in the MIRIAD
task IMMERGE.  We chose this combination method for several reasons,
specifically that it less sensitive than other methods to errors in
the model for the Parkes beam, is not overly computationally intensive
and produces reliable results.  The ATCA and Parkes data are
deconvolved separately, Fourier transformed, reweighted, linearly
combined and then inverse Fourier transformed.  If $V_{int}(k)$ is the
Fourier transform of the deconvolved ATCA mosaic and $V_{sd}(k)$ is
the Fourier transform of the deconvolved Parkes image, then the
Fourier transform of the combined image is given by
\begin{equation}
V_{comb}(k) = \omega^{\prime}(k) V_{int}(k) + f_{cal}\, \omega^{\prime\prime}(k)\,V_{sd}(k).
\end{equation}
The weighting functions, $\omega^{\prime}(k)$ and
$\omega^{\prime\prime}(k)$, are defined such that $\omega^{\prime}(k)
+ \omega^{\prime\prime}(k)$ is a Gaussian whose full-width half-max is
the same as the synthesized ATCA beam.  In this way the ATCA data are
downweighted on short baselines and the Parkes data are downweighted
on the longer baselines.  The calibration factor, $f_{cal}$, scales
the Parkes data to match the brightness temperature scale of the ATCA
data.  This factor is determined by comparing the ATCA and Parkes
datasets at every pixel and frequency in the range of overlapping
spatial frequencies, which for an ATCA mosaic and a Parkes image is
$\sim 120\lambda$ to $\sim 190\lambda$, where $\lambda = 21$ cm.  For
the SGPS we found that $f_{cal}$ was equivalent to one to within 2\%,
which adds confidence to the calibration of the individual Parkes and
ATCA datasets.  The final combined \HI\ cubes are regridded to
Galactic coordinates with a Cartesian projection.

\section{Resultant Data Products}
\label{sec:products}
There are three resultant \HI\ data products: the Parkes \HI\ cubes,
ATCA+Parkes continuum subtracted \HI\ cubes and ATCA+Parkes \HI\ cubes
with continuum emission.  All three data products are available for viewing and
downloading at the ATNF \HI\ Surveys Archive at
\url{http://www.atnf.csiro.au/research/HI/common}.

\subsection{Parkes Data Cubes}
\label{sec:pks_cubes}
\begin{figure}
\centering
\plotone{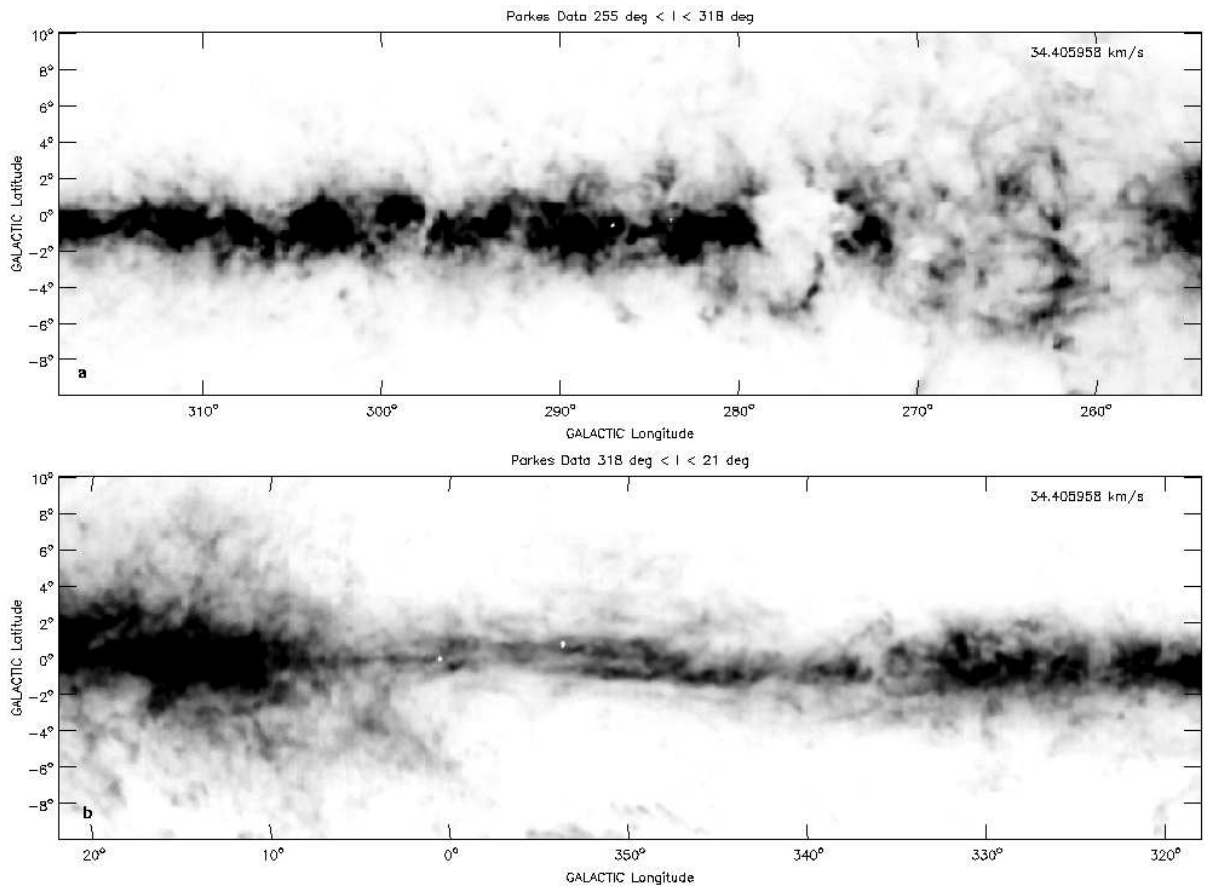}
\caption[]{Parkes data at $v=34.4$ \kms\ for the region $255\arcdeg
  \leq l \leq 318\arcdeg$ ({\bf a}) and  $318\arcdeg
  \leq l \leq 21\arcdeg$ ({\bf b}).  The greyscale is linear from 3 to
  100 K and has been intentionally saturated to bring out the
  low-level emission.  
\label{fig:pks1}}
\end{figure}

Because the Parkes survey covers a much larger area than the ATCA
survey these data are made available separately.  The Parkes data are
presented as twelve cubes, most covering $17\arcdeg \times 20\arcdeg$
($l \times b$).  Adjacent cubes overlap by about 4 degrees in
longitude to allow for convenient subsequent combination.  The center
position, size and sensitivity of the Parkes cubes are given in
Table~\ref{tab:pkscubes}.  The final cubes have an angular resolution
of 15\arcmin\ and a velocity resolution of $0.82$ \kms.  The rms noise
in line-free channels is approximately 180 mK for the regions $1\fdg5
\leq |b| \leq 10\arcdeg$, and approximately 60 mK in the region $|b|
\leq 1\fdg5$, which was observed many times.  However, stray radiation
from the far sidelobes can cause spurious emission with amplitudes as
high as $\sim 1$K, so low brightness temperature regions should be
viewed with caution.  As an example of the data quality we display the
entire Parkes dataset at a velocity of $v=34.4$ \kms\ in
Figures~\ref{fig:pks1}.  These images show the variety of size scales
represented in the Parkes data alone.

\subsection{Combined Parkes and ATCA Cubes}
\label{sec:images}
\begin{figure}
\centering
\includegraphics[angle=-90,width={0.9\textwidth}]{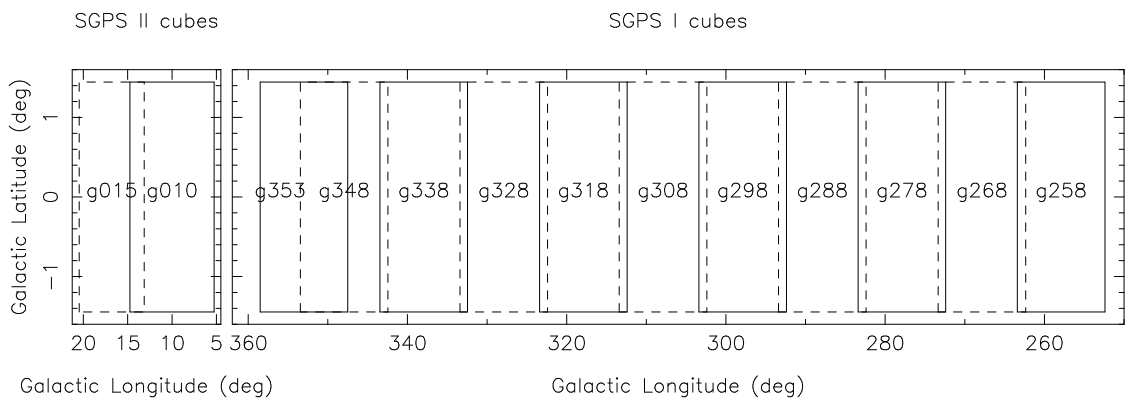}
\caption[]{Coverage of the SGPS I \& II continuum subtracted combined
  cubes.  Field names are the names used for the data cubes as
  released and as given in Table~\ref{tab:combcubes}.
\label{fig:cubepos}}
\end{figure}

A description and samples of the final combined Parkes and ATCA cubes
are presented here.  The data are available from the ATNF \HI\ Surveys
Archive. The cube sizes, center positions, line-free rms noise per
channel and synthesized beam sizes for the continuum subtracted data
and data containing continuum emission are given in Tables
\ref{tab:combcubes} and \ref{tab:contcubes}, respectively.  All cubes
have 40 arcsec pixels and a channel separation, $\delta v =0.82$ \kms.
The velocity range of the imaged cubes differs from cube to cube.
These ranges were chosen to include the bulk of the Galactic emission
and minimize cube size for download.  The mean rms in the line-free
channels is 1.6 K for both the SGPS I and II continuum subtracted
cubes.  The mean rms for the cubes containing continuum emission is
1.8 K.  The positions and dimensions of the combined continuum
subtracted cubes are shown in Figure \ref{fig:cubepos}.

\begin{figure}
\centering
\includegraphics[angle=-90,width={\textwidth}]{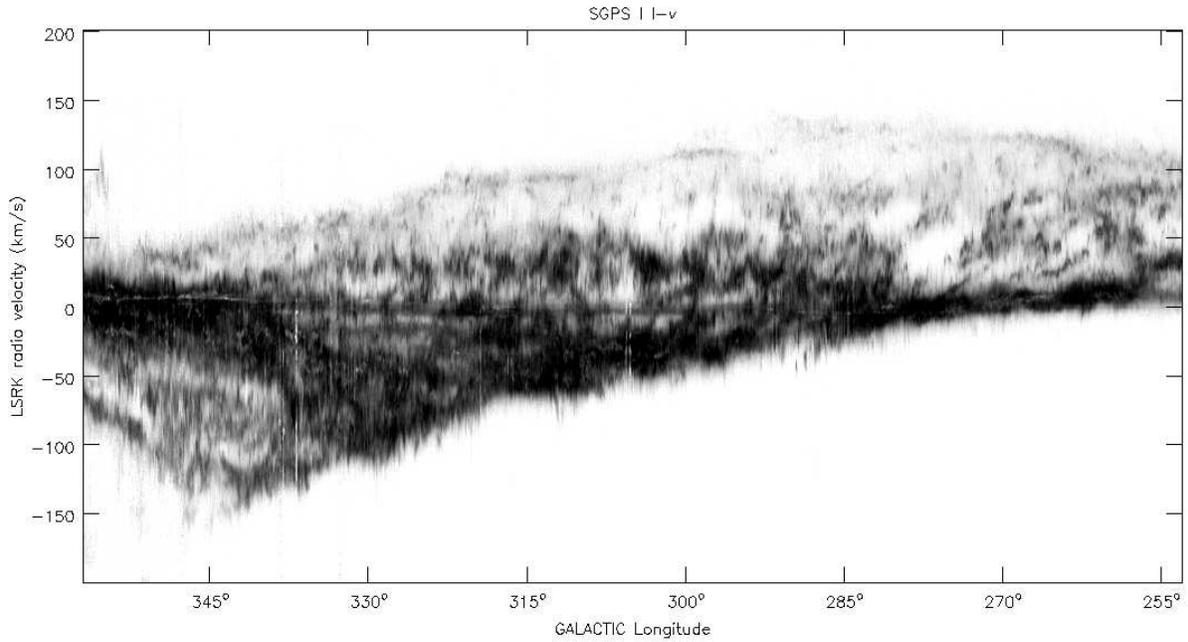}
\caption[]{Longitude-velocity diagram of continuum subtracted data for
  the SGPS I region at $b=0\arcdeg$.  The greyscale runs from 3 to 120
  K and uses a square-root transfer function.  The occasional white vertical
  stripes (e.g.\ near $l=306\arcdeg$) in the image are the result of
  \HI\ absorption towards strong continuum sources that have been
  subtracted.  Similarly, the horizontal stripes, especially near
  $v=0$ \kms\ are due to \HI\ self-absorption in the local gas.
\label{fig:sgps1}}
\end{figure}

\begin{figure}
\centering
\includegraphics[angle=-90,width={\textwidth}]{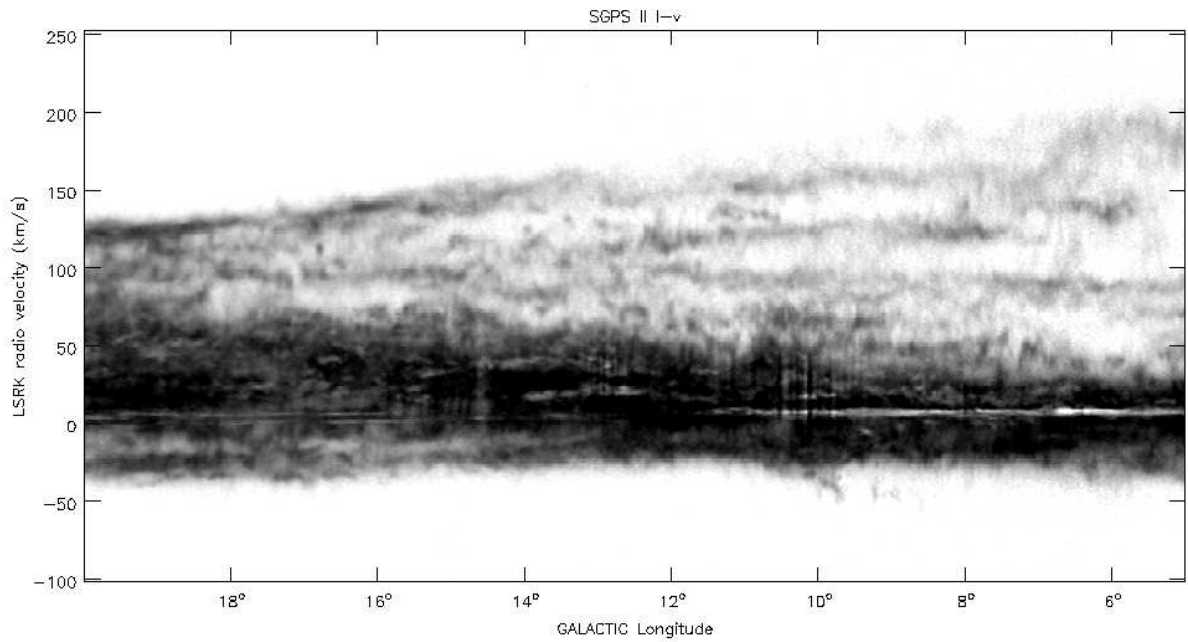}
\caption[]{Longitude-velocity diagram of continuum subtracted data for
  the SGPS II region at $b=0\arcdeg$.  The greyscale runs from 3 to
  120 K and uses a square-root transfer function.  As in
  Figure~\ref{fig:sgps1}, the vertical stripes seen here are due to
  continuum absorption and the horizontal stripes near $v=0$ \kms\ are
  due to \HI\ self-absorption.
\label{fig:sgps2}}
\end{figure}

\begin{figure}
\centering
\plotone{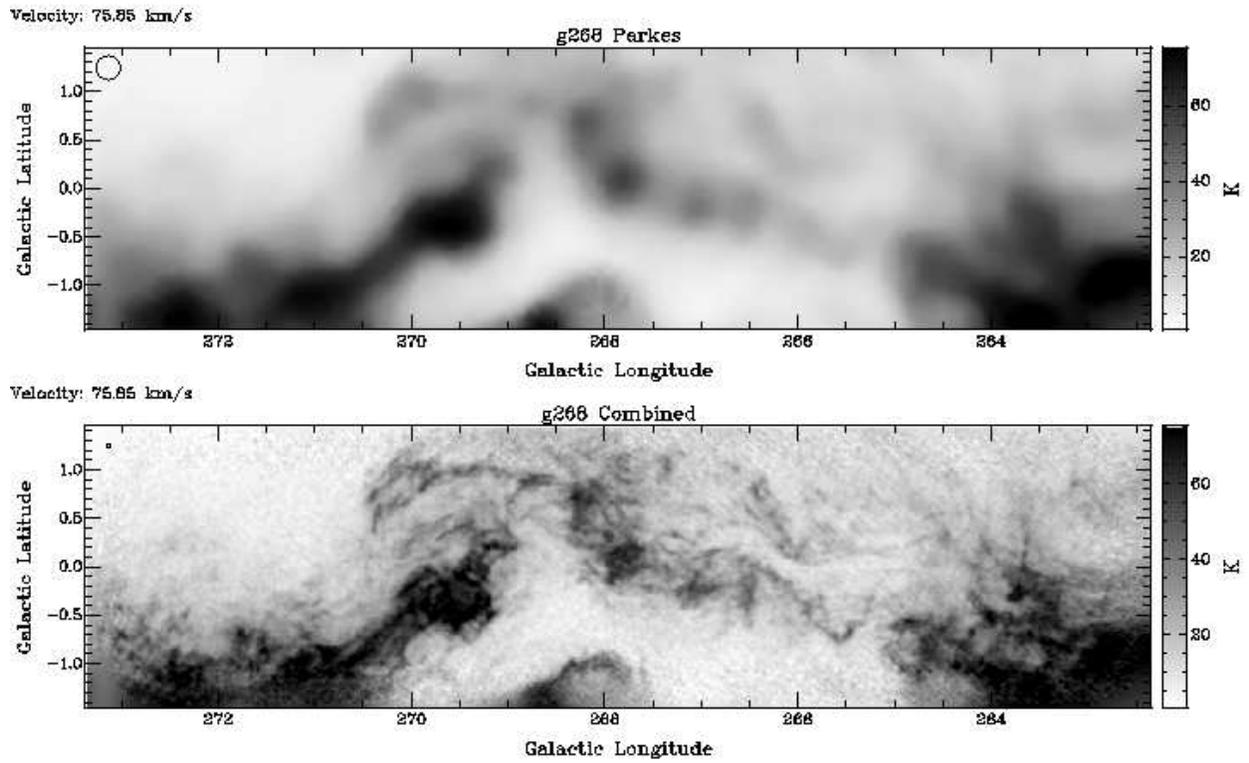}
\caption[]{Comparison of Parkes and Combined data for the g268 cube at
  $v=75.05$ \kms.  The beam size is marked as an open circle in the
  upper left-hand corner of each image.  The greyscale is the same for
  the two panels, running linearly from 1 to 75 K.
\label{fig:sgps_comp}}
\end{figure}

Because of the large volume of data in the SGPS it is impractical to
display all channels of the data cubes here.  Instead we have chosen
to present only longitude-velocity ({\em l-v}) diagrams at
$b=0\arcdeg$ and an example velocity channel from one of the cubes.
The {\em l-v} diagrams for SGPS I and II are shown in Figures
\ref{fig:sgps1} and \ref{fig:sgps2}, respectively.  There are
approximately 2900 independent spectra in the SGPS I {\em l-v} diagram
at $b=0\arcdeg$ and 300 in the SGPS II {\em l-v} diagram at
$b=0\arcdeg$.  Figure \ref{fig:sgps_comp} is a single velocity channel
at $v=75.05$ \kms\ from the continuum subtracted Parkes+ATCA g268
cube, compared with the same velocity channel in the Parkes data
alone.  This figure shows the dramatic increase in resolution of the
SGPS combined image compared with just the single dish data.
Figure~\ref{fig:sgps_comp} also demonstrates that the combined data
accurately represent the large scale structure present in the single
dish data.

\subsection{Data Quality and Artifacts}
\label{sec:quality}
We have taken care to reduce the number of artifacts present in the
SGPS data, however there are some artifacts that are unavoidable.
These are, in general, artifacts related to gain calibration of the
Parkes multibeam and limited {\em u-v} coverage in the ATCA data.  The
Parkes data have some residual longitudinal striping caused by
inter-beam gain variations in the multibeam and errors in the
estimation of the system temperature, $T_{\rm sys}$, as explained in
\S \ref{subsec:calibration}.  For the continuum subtracted cubes,
errors due to $T_{\rm sys}$ are subtracted, leaving only gain
variations.  These variations are least evident in the region $|b| \leq
1\fdg5$, which was observed many times, but are apparent in the higher
latitude data.  For the cubes containing continuum emission, errors in
the estimation of $T_{\rm sys}$ effect the reliability of brightness
temperature values on the largest angular scales.  Once again, this
problem is mitigated for data in the region $|b| \leq 1\fdg5$ where
ten independently calibrated datasets were averaged together.

The Parkes cubes also contain some artifacts related to saturation.
There are four positions along the Galactic plane where strong
continuum emission saturated the data samplers at Parkes.  At these
positions the observed fluxes are unreliable so we have blanked the
images.  These appear as white spots in Figure~\ref{fig:pks1}.

The ATCA data suffer from artifacts resulting from a limited number of
snapshots and incomplete {\em u-v} coverage.  Because the
interferometric data contain a small number (usually $\sim 40$) of
short duration hour-angle cuts these data have a dynamic range of
about 150:1.  For Galactic plane \HI\ emission the measured brightness
temperature is generally less than 150 K so the images are sensitivity
limited, not dynamic range limited.  However, towards the numerous
bright continuum sources in the Galactic plane, the dynamic range is
the fundamental limitation.  These continuum sources appear in
absorption in the continuum subtracted data where they can limit the
images as much as in the non-continuum subtracted data.  The limited
{\em u-v} coverage also leads to a number of radial and azimuthal
sidelobe artifacts around bright continuum sources that are not fully
removed by the deconvolution process.  The SGPS I data suffer from
effects related to the fact that all baselines are multiples of 15.3
m.  The result is a grating lobe of radius $0\fdg8$ around strong
continuum sources.  The addition of the Parkes data to the
interferometric images reduces the effect of the grating lobe but it
remains at a level of about 5\% of the continuum peak.  This artifact
is further reduced in the SGPS II cubes, which contain north-south
hybrid arrays with a wider variety of baseline spacings.

\section{Discussion}
\label{sec:discussion}
SGPS data can be applied to a number of scientific applications that
demand high-resolution \HI\ imaging of large areas. In the Galactic
ISM many structures are large and have complex interrelationships,
which are not apparent from more piecemeal observations.  It is in
these studies that the SGPS can excel, revealing fine detail that
would not have been observed in the large single dish surveys
previously available. However, a realistic 3-dimensional picture of
the ISM is only possible if the large-scale flux density is measured,
which is not measured in interferometric surveys. The large-scale flux
is essential for producing a true estimate of the \HI\ mass and
dynamics of the Galaxy and understanding the nature of ISM structures.
For example, there are some features which are principally
perturbations in velocity space caused by turbulence whereas there are
others that have massive density variations (e.g.\ shells with
voids).  Purely interferometric surveys, while showing details well,
can fail to distinguish between these two types of features.

This paper gives a description of the three data products of the SGPS
surveys of the \HI\ line: Parkes continuum subtracted \HI\ cubes with
a resolution of 15\arcmin, combined Parkes and ATCA continuum
subtracted \HI\ cubes with a resolution of $2\farcm2$ and combined
Parkes and ATCA cubes with a resolution of $\sim 1\farcm6$ containing
continuum emission.  Each of these data products was produced for use
in different general scientific purposes.  The Parkes \HI\ data,
though low resolution, cover an extended latitude range.  These data
are most useful for large-scale panoramic views of the \HI, tracing
Galactic structure and studies of discrete structures that extend
beyond the thin disk of the Galactic plane.  The combined Parkes and
ATCA continuum subtracted cubes provide high-resolution \HI\ images
with maximum surface brightness sensitivity.  These data are
best-suited for high-resolution examinations of \HI\ emission
structures.  The combination of the ATCA data with Parkes data ensures
that these cubes can be used to explore the ISM on multi-degree
scales.  Finally, the combined Parkes ATCA data containing continuum
emission provide accurately calibrated data at a slightly higher
angular resolution.  These data were produced specifically for use in
\HI\ absorption studies, which require good knowledge of the continuum
flux as well as the \HI\ flux in order to accurately measure optical
depth.

\section{Conclusions}
\label{sec:conclusions}
The Southern Galactic Plane Survey (SGPS) is a major new data resource
for Galactic \HI\ astronomy, providing an unprecedented view of \HI\
in the inner Galaxy in the southern hemisphere.  As part of the
International Galactic Plane Survey, the SGPS contributes to nearly
continuous coverage of the 1st, 2nd, and 4th quadrants of the plane of
the Galaxy.  We have presented the \HI\ data products which are now
publicly available via the ATNF \HI\ Surveys archive.  The \HI\ data
products include Parkes \HI\ cubes with 15 arcmin resolution and $\sim
100$ mK rms sensitivity covering the area $253\arcdeg \leq l \leq
20\arcdeg$ and $|b| \leq 10\arcdeg$, continuum subtracted combined
ATCA and Parkes \HI\ cubes with 2 arcmin resolution and $\sim 1.6$ K
rms sensitivity covering the regions $253\arcdeg \leq l \leq
358\arcdeg$, $|b| \leq 1\fdg5$ (SGPS I) and $5\arcdeg \leq l \leq
20\arcdeg$, $|b| \leq 1\fdg5$ (SGPS II), and combined ATCA and Parkes
\HI\ cubes of the same regions including the continuum emission.

\acknowledgements The ATCA and the Parkes Radio Telescope are part of
the Australia Telescope which is funded by the Commonwealth of
Australia for operation as a National Facility managed by CSIRO.  This
research was supported by NSF grants AST-9732695 and AST-0307603 to
the University of Minnesota and AST-0307358 to Harvard University.
Additional support was provided through a NASA Graduate Student
Researchers Program (GSRP) Fellowship to N.\ M.\ M.-G.\ while at the
University of Minnesota.  We would like to thank the staff of the
Australia Telescope National Facility for their support of this
project, especially M.\ Calabretta, R.\ Haynes, D.\ McConnell, J.\
Reynolds, R.\ Sault, R.\ Wark, and M.\ Wieringa.


\normalsize

\begin{deluxetable}{lccccc}
\tabletypesize{\footnotesize}
\rotate
\tablecaption{Basic parameters for the SGPS \HI\ survey}
\label{tab:params}
\tablewidth{\textheight}
\tablehead{
\colhead{Dataset} & \colhead{Angular Resolution} & \colhead{Spectral Resolution} &\colhead{Longitude
  coverage} & \colhead{Latitude coverage} & \colhead{1-$\sigma$ Sensitivity}
}
\startdata
Parkes SGPS & 15\arcmin & 0.8 \kms\ & $253\arcdeg \leq l \leq 20\arcdeg$ &$|b|
\leq 10\arcdeg $ & 180 mK  (60 mK\tablenotemark{a}) \\
ATCA + Parkes SGPS I & $2\farcm2$ & 0.8 \kms\ & $253\arcdeg \leq l \leq
358\arcdeg$ & $|b| \leq 1\fdg5$ & $\sim 1.6$ K\\
ATCA + Parkes SGPS II & $\sim 3\farcm3$ & 0.8 \kms\ & $5\arcdeg \leq l \leq
20\arcdeg$ & $|b| \leq 1\fdg5$ & $\sim 1.6$ K\\
\enddata
\tablenotetext{a}{The rms in the latitude range $|b| \leq 1\fdg5$.}
\end{deluxetable}

\begin{deluxetable}{lcccccccccccc}
\tabletypesize{\footnotesize}
\rotate
\tablewidth{\textheight}
\tablecaption{Cube properties for the Parkes \HI\ Cubes.  Column
  headings are as follows: Cube name; center Galactic longitude, $l$;
  center Galactic latitude, $b$; longitude pixel size, $\delta l$,
  latitude pixel size, $\delta b$; number of pixels in $l$, npix$_l$;
  number of pixels in $b$, npix$_b$; longitude coverage, $\Delta l$;
  latitude coverage, $\Delta b$; minimum velocity, $v_{min}$; maximum
  velocity, $v_{max}$; velocity channel width, $\delta v$; and cube
  rms.
\label{tab:pkscubes}}
\tablehead{
\colhead{cube} & \colhead{$l$} & \colhead{$b$} & \colhead{$\delta l$} & \colhead{npix$_l$}  & \colhead{$\delta b$} & \colhead{npix$_b$} & \colhead{$\Delta l$} &  \colhead{$\Delta b$} & \colhead{$v_{min}$} & \colhead{$v_{max}$} & \colhead{$\delta v$}  & \colhead{rms}\\ 
\colhead{} & \colhead{($\arcdeg$)} & \colhead{($\arcdeg$)}  & \colhead{($\arcmin$)} & \colhead{}&  \colhead{($\arcmin$)} & \colhead{} & \colhead{($\arcdeg$)} & \colhead{($\arcdeg$)} &  \colhead{(\kms)} & \colhead{(\kms)}  & \colhead{(\kms)}  &\colhead{(mK)}
}
\startdata
p268 & 264.5 & 0.0 & 4 & 345  & 4 & 305 & 23.00 & 20.33 & -149.8 & 220.4 & 0.82 &  130 \\
p278 & 277.5 & 0.0 & 4 & 255  & 4 & 305 & 17.00 & 20.33 & -149.8 & 250.0 & 0.82 &  130 \\
p288 & 287.8 & 0.0 & 4 & 255  & 4 & 305 & 17.00 & 20.33 & -180.3 & 250.0 & 0.82 &  140 \\
p298 & 297.5 & 0.0 & 4 & 220  & 4 & 300 & 14.67 & 20.0 & -200.1 & 199.7 & 0.82 &  200 \\
p308 & 307.5 & 0.0 & 4 & 220  & 4 & 300 & 14.67 & 20.0 & -200.1 & 199.7 & 0.82 &  210 \\
p318 & 317.5 & 0.0 & 4 & 225  & 4 & 300 & 15.00 & 20.0 & -350.2 & 349.8 & 0.82 &  200 \\
p328 & 327.5 & 0.0 & 4 & 225  & 4 & 300 & 15.00 & 20.0 & -200.1 & 250.0 & 0.82 &  200 \\
p338 & 337.5 & 0.0 & 4 & 220  & 4 & 300 & 14.67 & 20.0 & -299.9 & 300.3 & 0.82 &  230 \\
p348 & 347.5 & 0.0 & 4 & 225  & 4 & 300 & 15.00 & 20.0 & -299.9 & 199.7 & 0.82 &  190 \\
p358& 357.5 & 0.0 & 4 & 225  & 4 & 300 & 15.00 & 20.0 & -299.9 & 199.7 & 0.82 &  180 \\

p003 & 3.0 & 0.0 & 4 & 249  & 4 & 300 & 16.60 & 20.0 & -399.6 & 381.1 & 0.82 &  170 \\
p015 & 15.0 & 0.0 & 4 & 225  & 4 & 300 & 15.00 & 20.0 & -149.8 & 300.3 & 0.82 &  220 \\
\enddata
\end{deluxetable}

\begin{deluxetable}{lccccccccccccc}
\tabletypesize{\footnotesize}
\rotate
\tablewidth{\textheight}
\tablecaption{Properties of the ATCA and Parkes combined \HI\
  continuum subtracted cubes.  Column headings are as follows: Cube
  name; center Galactic longitude, $l$; center Galactic latitude, $b$;
  longitude pixel size, $\delta l$, latitude pixel size, $\delta b$; number of
  pixels in $l$, npix$_l$; number of pixels in $b$, npix$_b$;
  longitude coverage, $\Delta l$; latitude coverage, $\Delta b$;
  minimum velocity, $v_{min}$; maximum velocity, $v_{max}$; velocity
  channel width, $\delta v$; synthesized beam size, $\theta$; and the
  cube per channel rms.
\label{tab:combcubes}}
\tablehead{
\colhead{cube} & \colhead{$l$} & \colhead{$b$} & \colhead{$\delta l$} & \colhead{npix$_l$}  & \colhead{$\delta b$} & \colhead{npix$_b$} & \colhead{$\Delta l$} &  \colhead{$\Delta b$} & \colhead{$v_{min}$} & \colhead{$v_{max}$} & \colhead{$\delta v$} & \colhead{$\theta $}  & \colhead{rms}\\ 
\colhead{} & \colhead{($\arcdeg$)} & \colhead{($\arcdeg$)}  & \colhead{($\arcsec$)} & \colhead{}&  \colhead{($\arcsec$)} & \colhead{} & \colhead{($\arcdeg$)} & \colhead{($\arcdeg$)} &  \colhead{(\kms)} & \colhead{(\kms)}  & \colhead{(\kms)} & \colhead{($\arcsec \times \arcsec$)}  &\colhead{(K)}
}
\startdata
g258 & 257.9 & 0.0 & 40 & 990  & 40 & 260 & 11.00 & 2.89 & -112.1 & 216.8 & 0.82 & $ 130\times 130$  & 1.4 \\
g268 & 267.9 & 0.0 & 40 & 990  & 40 & 260 & 11.00 & 2.89 & -131.9 & 197.0 & 0.82 & $ 130 \times 130$   & 1.4 \\
g278 & 277.9 & 0.0 & 40 & 990  & 40 & 260 & 11.00 & 2.89 & -99.8 & 229.2 & 0.82 & $ 130 \times 130$  & 1.3 \\
g288 & 287.9 & 0.0 & 40 & 990  & 40 & 260 & 11.00 & 2.89 & -112.1 & 216.8 & 0.82 & $ 130 \times 130$  & 1.6 \\
g298 & 297.9 & 0.0 & 40 & 990  & 40 & 260 & 11.00 & 2.89 & -132.7 & 196.2 & 0.82 & $ 130 \times 130$  & 1.4 \\
g308 & 307.9 & 0.0 & 40 & 990  & 40 & 260 & 11.00 & 2.89 & -145.1 & 183.9 & 0.82 & $ 130 \times 130$  & 1.5 \\
g318 & 317.9 & 0.0 & 40 & 990  & 40 & 260 & 11.00 & 2.89 & -161.6 & 167.4 & 0.82 & $ 130 \times 130$  & 1.6 \\
g328 & 327.9 & 0.0 & 40 & 990  & 40 & 260 & 11.00 & 2.89 & -186.3 & 142.6 & 0.82 & $ 130 \times 130$  & 1.7 \\
g338 & 337.9 & 0.0 & 40 & 990  & 40 & 260 & 11.00 & 2.89 & -202.8 & 126.1 & 0.82 & $ 130 \times 130$  & 1.9 \\
g348 & 347.9 & 0.0 & 40 & 990  & 40 & 260 & 11.00 & 2.89 & -247.3 & 164.9 & 0.82 & $ 130 \times 130$  & 1.9 \\
g353 & 353.0 & 0.0 & 40 & 990  & 40 & 260 & 11.00 & 2.89 & -300.1 & 150.0 & 0.82 & $ 145 \times 90$  & 2.6 \\
g010 & 10.0 & 0.0 & 40 & 853  & 40 & 260 & 9.48 & 2.89 & -145.1 & 266.3 & 0.82 & $ 160 \times 110$  & 1.8 \\
g015 & 16.8 & 0.0 & 40 & 660  & 40 & 260 & 7.33 & 2.89 & -128.6 & 200.3 & 0.82 & $ 200 \times 130$  & 1.4 \\
\enddata
\end{deluxetable}

\begin{deluxetable}{lccccccccccccc}
\tabletypesize{\footnotesize}
\rotate
\tablewidth{\textheight}
\tablecaption{Properties of the ATCA and Parkes combined \HI\
  cubes containing continuum emission.  Column headings are the same
  as in Table~\ref{tab:combcubes}.  
\label{tab:contcubes}}
\tablehead{
\colhead{cube} & \colhead{$l$} & \colhead{$b$} & \colhead{$\delta l$} & \colhead{npix$_l$}  & \colhead{$\delta b$} & \colhead{npix$_b$} & \colhead{$\Delta l$} &  \colhead{$\Delta b$} & \colhead{$v_{min}$} & \colhead{$v_{max}$} & \colhead{$\delta v$} & \colhead{$\theta $}  & \colhead{rms}\\ 
\colhead{} & \colhead{($\arcdeg$)} & \colhead{($\arcdeg$)}  & \colhead{($\arcsec$)} & \colhead{}&  \colhead{($\arcsec$)} & \colhead{} & \colhead{($\arcdeg$)} & \colhead{($\arcdeg$)} &  \colhead{(\kms)} & \colhead{(\kms)}  & \colhead{(\kms)} & \colhead{($\arcsec \times \arcsec$)}  &\colhead{(K)}
}
\startdata
g255 & 255.4  & 0.0  &-0   &  721   &  30   &   351   &  -6.0 &  2.9  &  -227.5 & 265.5 &  0.82 &  $143 \times 110$  & 1.2\\
g260 & 260.4  & 0.0 & 30 &    721 &    30 &     351  &   -6.0 &  2.9  &  -227.5  &265.5 &  0.82 &  $124 \times 120$  &    1.4\\
g265 & 265.5 & 0.0 & 30 &    721  &   30  &    351   &  -6.0 &  2.9& -227.5  &265.5  & 0.82  & $138 \times 110$  &      1.8\\
g270 &270.4 & 0.0 & 30 &    721 &    30 &     361 &    -6.0 &  3.0   & -227.5 & 265.5 &  0.82 &  $132 \times 114$ &      1.1\\
g275 &275.4 &  0.0 & 30 &    721 &    30 &     361 &    -6.0 &  3.0   & -227.5 & 265.5 &  0.82 &  $131 \times 116$ &       1.2\\
g280 &280.4 &  0.0 & 30 &    721 &    30 &     346 &    6.0 &  2.9 &   -227.5 & 265.5 &  0.82 &  $133 \times 112$ &      1.2\\
g285 &285.4 &  0.0 & 30 &    721 &    30 &    351   &  6.0 &  2.9 & -227.5 & 265.5  & 0.82 &  $126 \times 115$ &  1.8 \\ 
g290 &290.4 &  0.0 & 30 &    721 &    30 &    351 &    6.0 &  2.9 &-227.5 & 265.5 &  0.82 &  $129 \times 112$&  1.5\\ 
g295 &295.4 &  0.0 & 30 &    721 &    30 &     351 &    6.0 &  2.9 &   -227.5 & 265.5 &  0.82 &  $123 \times 116$  &  1.4\\
g300 & 300.4 &  0.0 & 30 &    721 &    30 &     351 &    6.0 &  2.9 &   -227.5 & 265.5 &  0.82 &  $124 \times 113$ & 1.6\\
g305 & 305.4 &  0.0 & 30   &  721   &  30   &   351 &    6.0 &  2.9&   -227.5 & 265.5 &  0.82 &  $135 \times 112$  & 1.3\\
g310 & 310.4 &  0.0 & 30 &    717 &    30 &     348 &    6.0 &  2.9& -227.5 &  265.5 &  0.82 &  $131 \times 119$ & 1.8\\
g315 & 315.4 &  0.0 & 30 &   716 &    30 &     346 &    6.0 &  2.9 &   -227.5 & 265.5 &  0.82 &  $131 \times 116$ & 1.7\\
g320 & 320.4 &  0.0 & 30 &    721 &    30 &     346 &    6.0 &  2.9 &   -227.5 & 265.5 &  0.82 &  $129 \times 119$ & 1.7 \\
g325 & 325.4 &  0.0 & 30 &    721 &    30 &     351 &    6.0 &  2.9 &   -227.5 & 265.5 &  0.82 &  $142 \times 117$ & 1.7   \\
g330 & 330.4 &  0.0 & 30 &    721 &    30 &     356 &    6.0 &  3.0 &   -227.5 & 265.5 &  0.82 &  $129 \times 115$ & 2.2\\
g335 & 335.4 &  0.0 & 30 &    716 &    30 &     351 &    6.0 &  2.9 &   -227.5 & 265.5 &  0.82 &  $137 \times 126$ & 2.2 \\
g340 & 340.4 &  0.0 & 30 &    716 &    30 &     351 &    6.0 &  2.9 &   -227.5 & 265.5 &  0.82 &  $127 \times 121$ & 2.1\\
g345 & 345.4 &  0.0 & 30 &    716 &    30 &     346 &    6.0 &  2.9 &   -227.5 & 265.5 &  0.82 &  $147 \times 114$ & 2.6\\
g350 & 350.4 &  0.0 & 30 &    716 &    30 &     351 &    6.0 &  2.4 & -227.5 & 265.5 &  0.82 &  $152 \times 108$ & 2.4\\ 
g355 & 355.6 &  0.0 & 30 &    761 &    30 &     351 &    6.3 &  2.9 & -227.5 & 265.5 &  0.82 &  $161 \times 108$ &  3.4\\
g010 & 9.5 &  0.0 & 30 &    1200 &    30 &     350 &    10.0 &  2.9 & -227.5 & 265.5 &  0.82 &  $190 \times 93$ &  2.0\\
g015 & 16.5 &  0.0 & 30 &    860 &    30 &     350 &    7.2 &  2.9 & -227.5 & 265.5 &  0.82 &  $210 \times 92$ &  1.9\\
\enddata
\end{deluxetable}

\clearpage

\end{document}